# Iwahashi Zenbei's Sunspot Drawings in 1793 in Japan

Hisashi Hayakawa (1, 2)*, Kiyomi Iwahashi (3), Harufumi Tamazawa (4), Shin Toriumi (5, 6), Kazunari Shibata (4, 7)

(1) Graduate School of Letters, Osaka University, 5600043, Toyonaka, Japan.
(2) Research Fellow of Japan Society for the Promotion of Science, 1020083, Tokyo, Japan.
(3) National Institute of Japanese Literature, 1900014, Tachikawa, Japan.
(4) Kwasan Observatory, Kyoto University, 6078471, Kyoto, Japan.
(5) National Astronomical Observatory of Japan, 1818588, Mitaka, Japan.
(6) NAOJ Fellow, 1818588, Mitaka, Japan.
(7) The Unit of Synergetic Studies for Space, Kyoto University, 6068502, Kyoto, Japan.

*email: hayakawa@kwasan.kyoto-u.ac.jp

**Abstract**

Three Japanese sunspot drawings associated with Iwahashi Zenbei (1756-1811) are shown here from contemporary manuscripts and woodprint documents with the relevant texts. We revealed the observational date of one of the drawings to be 26 August 1793, and the overall observations lasted for over a year. Moreover, we identified the observational site for the dated drawing at Fushimi in Japan. We then compared his observations with group sunspot number and raw group count from Sunspot Index and Long-term Solar Observations (SILSO) to reveal its data context, and concluded that these drawings filled the gaps in understanding due to the fragmental sunspots observations around 1793. These drawings are important as a clue to evaluate astronomical knowledge of contemporary Japan in the late 19$^{th}$ century and are valuable as a non-European observation, considering that most sunspot observations up to mid-19$^{th}$ century are from Europe.

**1. Introduction**

One of the longest ongoing scientific research that has generated large datasets for the review of solar activity is sunspot counting (Owens, 2013). While telescopic observations have been carried out for over 150 years for observing solar flare since the Carrington event in 1859 (Carrington, 1859; Kimball, 1960; Tsurutani *et al.*, 2003; Cliver *et al.*, 2004; Hayakawa *et al.*, 2016b), telescopic observations for sunspots exist for over 400 years since Galilei's sunspot drawings forming one of the most important indices for solar physics (Galilei, 1613; Owens, 2013). These datasets contributed to the reconstruction of the Wolf number (Zürich number) by R. Wolf (*e.g.*, Waldmeier, 1961), and the group sunspot number by Hoyt and Schatten (1998). Recently, several authors revisited sunspot number ($S_N$) in general to make crucial contributions (Clette *et al.*, 2014; Svalgaard and Schatten, 2016; Vaquero *et al.*, 2016) based on the latest discussions on raw sunspot data within early modern scientific documents, consisting of sunspot number counting and sunspot drawings (*e.g.* Vaquero, 2007; Vaquero and Vázquez, 2009; Arlt, 2008, 2011; Diercke *et al.*, 2014; Usoskin *et al.*, 2015; Arlt *et al.*, 2016; Carrasco and Vaquero, 2016). Within these datasets, sunspot drawings are of greater value as they provide information not only on $S_N$ but also on sunspot area, distribution, locations, and so forth (Vaquero, 2007; Vaquero and Vázquez, 2009).

While most of these sunspot observations are from European astronomers (Vaquero, 2007), recent studies rediscovered sunspot drawings of other areas from non-European astronomers. They are of great importance to recover solar observations before the mid-19[th] century (Domínguez-Castro *et al.*, 2017; Denig and McVaugh, 2017). In this context, it must be noted that considerably longer traditions of sunspot observations are present not only in Europe (e.g., Stephenson and Willis, 1999; Willis and Stephenson, 2001) but also in East Asia or West Asia, even in pre-telescopic ages (Keimatsu, 1970; Clark and Stephenson, 1978; Willis *et al.*, 1980, 1996; Yau and Stephenson, 1988; Xu *et al.*, 2000; Willis and Stephenson, 2002; Lee *et al.*, 2004; Hayakawa *et al.*, 2015, 2017a, 2017b, 2017c; Tamazawa *et al.*, 2017a; Goldstein, 1969; Vaquero and Gallego, 2002). After the 17[th] century, these non-European astronomers began to make contact with European astronomy and some started to adopt European technology. In particular, contemporary Japanese astronomers imported European Astronomy and the associated technologies through trade with Dutch merchants, in order to make some sunspot observations (Watanabe, 1987). More significantly, Kunitomo Ikkansai (国友一貫斎) improved European telescopes and carried out continuous sunspot observation in 1835/1836 (Kanda, 1932; Yamamoto, 1937; Kubota and Suzuki, 2003). However, Tamazawa *et al.* (2017b) reported that one of the earliest sky watching events with telescopes occurred in 1793. It was Iwahashi Zenbei (岩橋善兵衛, 1756-1811) who operated the telescope that he constructed for sunspot observation in the sky-watching event. As participants left three accounts with sunspot drawings, we show these drawings and relevant records to find the dates of observation and analyze their descriptions.

## 2. Method

### 2.1. Analyses

In order to examine Iwahashi Zenbei's sunspot observations from the year 1793, we examined three documents from a contemporary source as explained below and compared their descriptions. We then analyzed the sunspot records to count the number of sunspots observed. Furthermore, we compared the data context in terms of group sunspot number and raw sunspot group counts to those by contemporary European sunspot observations. We have translated *kokuten* (黒點), which is a technical term for sunspot, as its literal meaning, "black spot," for historical significance, and as "sunspot" in the context of scientific discussions.

### 2.2. Source Documents

In order to fulfill the said purpose, we examined three historical documents written by participants in the sky-watching party in 1793. Listed below are the references to these documents in the storing archive using their abbreviations, authors, and reference numbers as follows:

BK-J: Tachibana Nankei (橘南谿), *Bouenkyo Kanshoyoki* (望遠鏡観諸曜記), MS MB-51-Ta in the Library of the National Astronomical Observatory of Japan [a manuscript in Japanese]

BK-N: Tachibana Nankei (橘南谿), *Bouenkyo Kanshoyoki* (望遠鏡観諸曜記), MS 463 in the Library of the International Research Center for Japanese Studies [a manuscript in Japanese]

BK-T: Tachibana Nankei (橘南谿), *Bouenkyo Kanshoyoki* (望遠鏡観諸曜記), MS Inada 44-210 in the Tsu City Library [a manuscript in Japanese]

KJ: Ban Koukei (伴蒿蹊), *Kanden Jihitsu* (閑田次筆), MS MY1491-2 in the National Institute for Japanese Literature [a woodprint in Japanese]

HZ: Iwahashi Zenbei (岩橋善兵衛), *Heitengi Zukai* (平天儀圖解), MS MY-1440-2 in the National Institute for Japanese Literature [a woodprint in Japanese]

*Bouenkyo Kanshoyoki* (BK) is a record of the earliest sky watching with telescopes in Japan held on 26 August 1793 at Fushimi, written in classic Chinese by Tachibana Nankei (1753-1805), the host of the sky watching. On the said day, Tachibana Nankei, who was a doctor, and Iwahashi Zenbei held a sky watching with 12 participants from various academic backgrounds. The participants observed not only the Sun, but also the Moon, Jupiter, Saturn, and Venus using Iwahashi Zenbei's telescope (Tamazawa *et al*., 2017b). At the time of writing, three variants of BK are known with exact location: BK-J, BK-N, and BK-T, according to the Union Catalogue of Early Japanese

Books[1]. Unfortunately, neither of them have explicit dating of copying in their colophons. All the variants commonly have observational records in 1793 (BK-N: 2b; BK-T: 2b; BK-J: 4a). BK-T and BK-J have additional description of another observation in 1795 (BK-T: f.4b; BK-J: 4a). BK-J has additional statement on the contemporary European astronomy with the name of Agawa Biren (阿川美廉) that does not appear in any variants. Therefore, chronologically we regard BK-N as a first edition or its copy, BK-T as a second edition or its copy, and BK-J as a copy by Agawa Biren. Nevertheless, we have to admit that we cannot determine if either BK-J or BK-T is the autograph manuscript by Tachibana Nankei himself in the current stage. Therefore, we show the sunspot drawing in each variant.

*Kanden Jihitsu* (KJ) is an essay with four volumes by Ban Koukei (1733-1806) published as a woodprint edition in 1806. This essay consists of miscellaneous topics, including the description of the sky watching in 1793 in classic Japanese. At the time of writing, we have at least 43 copies with exact location in various archives in Japan and the United States, according to the Union Catalogue of Early Japanese Books. We use a copy stored in the National Institute for Japanese Literature (NIJL) in our article.

*Heitengi Zukai* (HZ) is an introductory account for astronomy by Iwahashi Zenbei published as a woodprint edition in 1802. Iwahashi had studied Confucianism under Minagawa Kien at Kyoto, and hence the latter offered the introduction to the former's account. This account consists of explanations on various topics on astronomy and meteorology, with figures of observations of astronomical bodies and phenomena. In particular, this account aims at explaining how to use the *heitengi* (平天儀), a kind of star chart connecting disks of the Earth, the Moon, the Sun, and constellations of 28 lunar mansions, with basic knowledge of astronomy and observational records of the Sun, Moon, or planets with the telescopes he invented. At the time of writing, we have at least 38 known copies with exact location in various archives in Japan and Germany, according to the Union Catalogue of Early Japanese Books. We use a copy stored in the NIJL in our article.

### 3. Results and Discussions
### 3.1. Sunspot Observations
As documented in the previous section, we examined three contemporary sunspot records in BK, KJ, and HZ. BK and KJ have sunspot drawings (BK-N, f.3b; BK-T, f.3b; BK-J, f.4b; KJ, v.1 f.4a) named *nisshinsho* (日眞象), as shown in Figures 1a-c and Figure 2. HZ has sunspot drawing

---
[1] http://base1.nijl.ac.jp/infolib/meta_pub/G0001401KTG (last accessed on 10 Oct. 2017)

(f.35a) named *taiyouzu* (太陽圖), as shown in Figure 3. Here, we show their abbreviations, references, transcriptions, and translations below.

A: BK-N, f.1a, BK-T, f.1a, BK-J, f.1b

Transcription:

觀日，四邊有気如毛，気皆左旋，日面有黒點五，大小不一，善兵衛言，黒点歷十餘日，徑日面，若冬春之際，則黒點最多，又或見梵字形者，其色亦[2]眞黒，不能辧其何物

Translation:

Observations state that there are vapors similar to hair. All vapors rotate leftward. On the surface of the sun, there are five black spots. Their sizes were different from one another. Zenbei stated that these black spots go around across the solar disk spending more than ten days. The black spots moved across the solar surface. In the winter to spring, the number of black spots is the largest. We observed a sunspot whose shape resembled that of a Sanskrit character. The color of it was also deep black, and we are not able to know what it is.

B: KJ, v.1, f.1b

Transcription:

日面黒點五つ有り．大小一ならず．善兵衛言う．黒點十餘日を歷ると．日面に亙る．冬・春の間ハ，則ち黒點最多しと．又或ハ蚯蚓のごとく梵字のごとく形もつものあり．其色純黒ふて，それ何ものといふこと無し．

Translation:

There were five black spots on the surface of the Sun. Their sizes were different from one another. Zenbei stated that these black spots go around for more than ten days across the solar surface. From winter to spring, the number of sunspots is the largest. We have black spots whose shape was similar to that of an earthworm or a Sanskrit character. The color of it was deep black, and we cannot say what it is.

C: HZ, f.35a

Transcription:

---

[2] In BK-J (f.1b), it is written "赤 (red)" instead of "亦 (also)", although it seems a miscopying considering the context. The copyist of BK-J frequently miscopies this character in other places as well (e.g. BK-J, f.2a).

又図の如く日輪の中に黒きもの出る事あり形龍のごとく或ハ虫の如く大小数定りなし出る時は東の方より方より出て凡そ日数十四・五日計り有之西の方ニ終る

Translation:

In addition, as shown in the figure, sometimes there appear black objects in the solar disk. They resemble a dragon or a worm in terms of their shape. They are different in size. When they appear, they move from the east side to the west side over ~14–15 days.

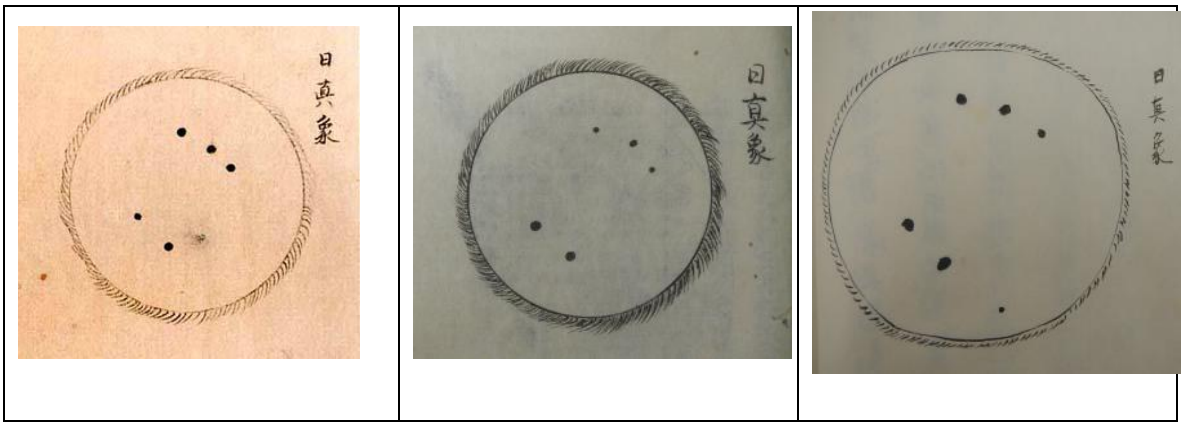

Figure 1: Sunspot drawing in variants of BK; (a) A variant in BK-N (courtesy: the National Astronomical Observatory of Japan); (b) A variant in BK-T (courtesy: Tsu City Library); (c) A variant in BK-J (courtesy: the International Research Center for Japanese Studies)

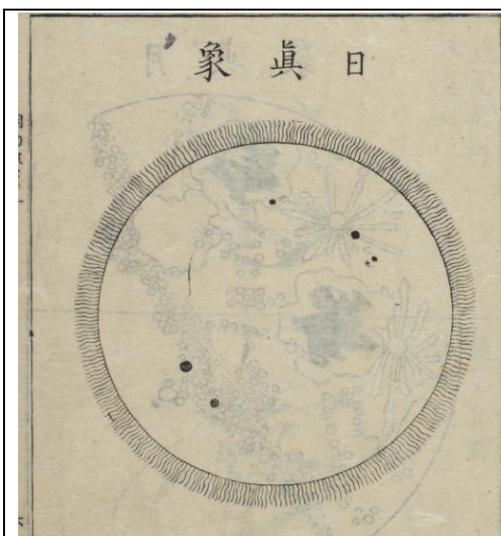

Figure 2: Sunspot drawing in KJ (courtesy: NIJL)

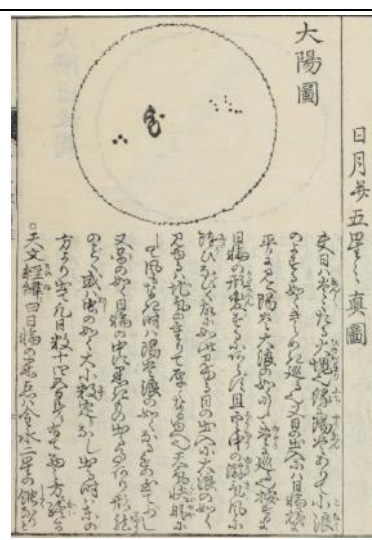

Figure 3: Sunspot drawing in HZ (courtesy: NIJL)

**3.2. Their Observational Dates and Sunspot Number**

With these three kinds of figures, we can easily distinguish the figure in HZ from those in BK and KJ. BK clearly dates its observation on Kansei 5$^{th}$ year, 7$^{th}$ month, and 20$^{th}$ day. According to the conversion table by Uchida (1992), we can convert this date in the Japanese lunisolar calendar to 26 August 1793 in the Gregorian calendar. In this observation, BK reports "five black spots" on "the solar surface" (BK-J, f.1a; BK-T, f.1a; BK-N, f.1b). We find the relevant text in KJ mostly Japanese translation of BK with some minor changes. The sunspot locations are generally identical within sunspot drawings of KJ and variants of BK except for BK-N (a copy by Agawa Biren): three dots at the upper right and two dots at the lower left. The sizes of sunspots found here are different from one another as documented in the relevant text. Note that, nevertheless, the sizes are differently depicted even within the variants of BK. Thus, we consider their dots for sunspot only represent their locations.

We therefore can fairly locate these sunspot drawings in BK and KJ on 26 August 1793, in the sky watching at Tachibana Nankei's personal residence (Kobayashi, 2009). Takayama Hikokurou (高山彦九郎, 1747-1793) visited his residence near "the mansion of lord of Satsuma (薩摩屋敷: N34°56′17″, E135°45′30″)", going southward from "Kujou-mura Takeda (久条村竹田)" on 19 May 1791 (THN, v.4, p.99). We also found his address as "Bungo Toyohashi-suji Tachiuri-cho (豊後豊橋筋立売町: N34°55′53″, E135°46′07″)" in the register of Iwahashi Zenbei in February 1798 (SNH, f.26a). While we cannot identify his residence in 1793, we can fairly locate his residence and hence the observational site in the midst of Fushimi.

While the two sunspot drawings in BK and KJ are identical with one another and their own descriptions, the sunspot drawing on HZ is different in its sunspot distribution, except for the variant of BK-N, which is considered a copy by Agawa Biren. HZ does not clarify the observational date. We find nine small sunspots on the right side of the solar disk, three medium sunspots in the left side of the solar disk, and a large and complex sunspot in the disk center. The largest one might be identical with a sunspot like "a dragon or a worm." Their sunspot number is hardly five as reported in the former two accounts, and hence considered the observational drawing from a different day. BK and KJ cited Iwahashi Zenbei's explanation that they have seen a sunspot that resembles "a Sanskrit character" (both in BK and KJ), or "an earthworm" (only in KJ), and hence these sunspots with strange structure are considered observed before the sky watching. Especially the latter seems to correspond to another report with similar expression to compare sunspot with "a dragon or worm" found in HZ, while we need to expect exaggeration in some extent in its size and shape. Therefore, we can regard the sunspot drawing in HZ as the drawing prior to the observation on 26 August 1793. Iwahashi Zenbei lived at Kaidzuka in Izumi (BK-J, f.1a; BK-T, f.1a; BK-N, f.1a; Kobayashi, 2009), and hence we identify his daily observational site around his residence (N34°27′, E135°21′), except the location for the sky-watching event on 26 August 1793.

Therefore, we regard the descriptions in BK and KJ as identical with their sunspot drawings in these documents, but we cannot date the sunspot drawing in HZ due to a lack of explicit dating. This fact attests that Iwahashi Zenbei had observed the solar surface for a considerable period. At least, he correctly understood that it takes 14–15 days for sunspots (black spots) to move from east to west in the solar disk. On the observational period, any source documents show no explicit value for the number of years he made the observations is given. Therefore, it is difficult to determine if Iwahashi Zenbei mentioned this trend as a general law of sunspot activity induced from his observational experience for multiple years or simply as a fact that he saw more sunspots in the previous winter and spring. Grammatically, it is likely Iwahashi Zenbei regarded this as a general law, as KJ in classic Japanese use the Present Tense to describe the sunspot rotation period and more appearance of sunspots in winter and spring. In the case it was in the context of explanation of his observational report, it is expected to see the Past Tense here. Note that it was in a declining phase of solar activity as noted later, and it might also possible to expect Iwahashi Zenbei to have witnessed much more sunspots in preceding spring and winter than in 26 August 1793. At least, Japanese astronomers in succeeding generation seem to interpret this as "a general law." For example, Kunitomo Ikkansai, who observed sunspots in 1835/6, owned a copy of KJ and is considered to examine this "general law" by himself (Nakamura, 2003). A similar apparent tendency is found in the Plate XXXI for the distribution of "solar spotted area" during 1832-1868 compiled by De La Rue *et al*. (1870) as well. While we do not find such periodicity within the Sun, we might be able to relate this apparent tendency with better atmospheric conditions for sunspot observations in winter as discussed in Willis *et al*. (1980).

### 3.3. The Telescope in Use

In HZ, the telescope used by Iwahashi Zenbei can be seen. It was an octagonal reflecting telescope called *kitenkyo* (窺天鏡, Figure 4) by Iwahashi Zenbei (HZ: f.34b). It was recorded that its perimeter was 25 cm and its length was 2.5 m (BK: f.1a; Kobayashi, 2009) and hence it is considered its diameter was approximately 8 cm. Iwahashi Zenbei was originally a craftsman who specialized in glasses, but he applied his knowledge and technique on the imitation of telescopes newly brought in from the Netherlands. He used an objective lens, erecting lens, and eyepiece lens in his telescope, and supported them with frames made of wood or paper, and fixed them using brass fittings (Date, 1933; Arisaka, 1952; Kobayashi, 2009). His telescopes became a standard in contemporary Japan, and hence used by the feudal lords such as those in Kii, Hikone, and Akashi, and scholars in urban areas such as Edo, Osaka, or Kyoto. In addition, Ino Tadataka (伊能忠敬) used his telescope to make the earliest measured map of Japan (Watanabe, 1987). While Iwahashi Zenbei himself did not clarify how he filtered the sunlight during solar observations, we might estimate the usage of *zongurasu*; a kind of filtering glass originally known as *zonglas* that was imported from the Netherland around mid-18$^{th}$ century (*Tokugawa Jikki*, v.9, p.294; Vos, 2014; Zuidelvaart, 2007).

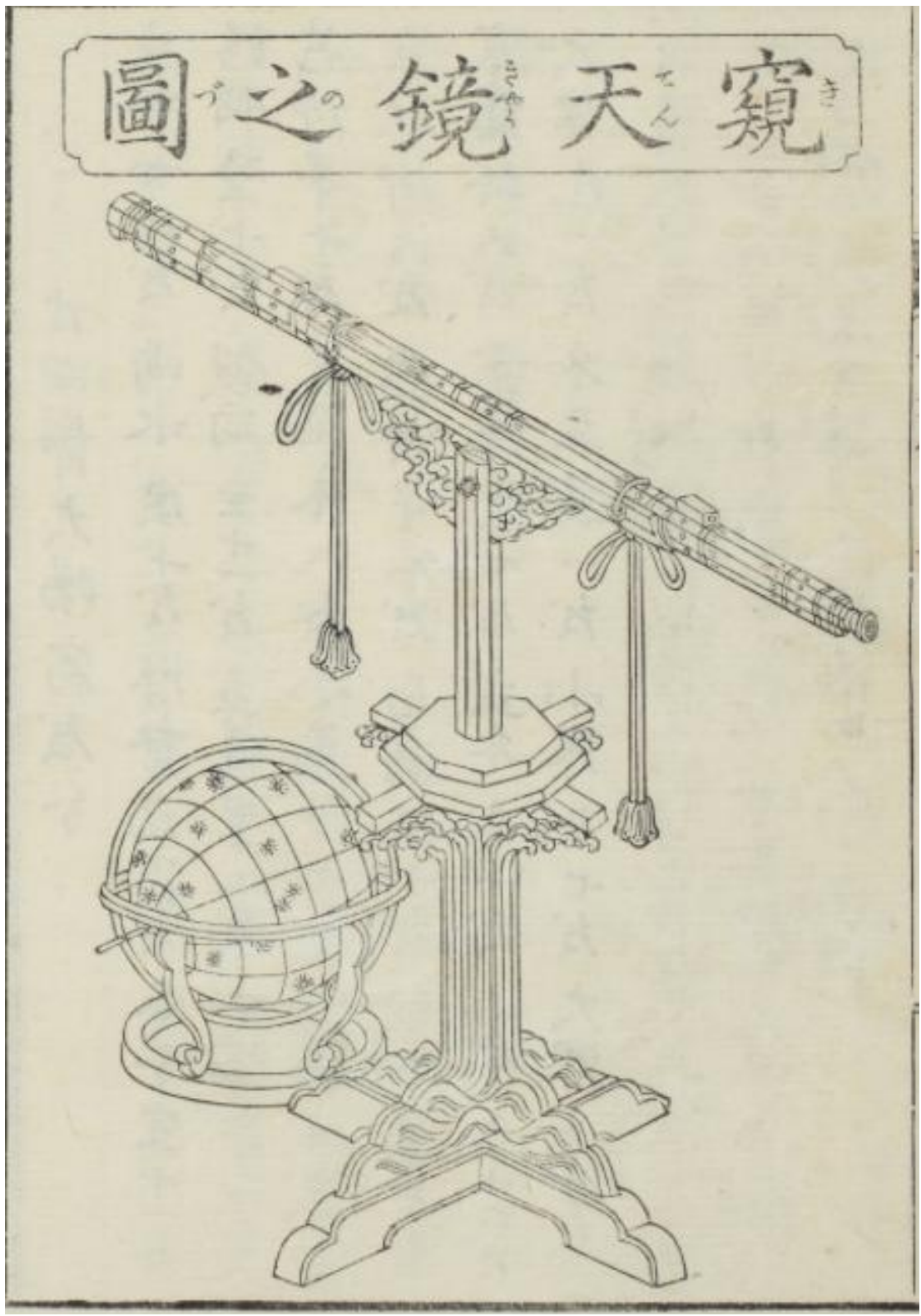

Figure 4: *kitenkyo* (窺天鏡) by Iwahashi Zenbei (HZ: f.34b).

## 3.4. Data Context

Figures 3a and 3b show the location of the sunspot drawing dated 26 August 1793. Figures 5a, 5b, and 5c show the annual sunspot number, the monthly sunspot number spanning from 1790 to 1799, as reported by Clette *et al*. (2014), and the raw group count reported in Vaquero et al. (2016) in comparison with this study. Table 1 shows the data context of Iwahashi Zenbei's sunspot drawing in comparison with the daily raw group count shown in Vaquero *et al*. (2016). As shown in Figures 3a and 3b, this observation is situated in the declining phase of solar activity from the nearest maximum in 1787 to the Dalton Minimum (Usoskin, 2013; Clette *et al*., 2014).

Nevertheless, as shown in Table 1 and Figure 5c, there are only 18 previously known sunspot observations in 1793. We have observations by Bode at Berlin, Schroter at Lilienthal, Huber at Basel, Staudacher at Nuremberg, and Hahn at Basel in this period (*e.g.* Arlt, 2008; Vaquero *et al*., 2016). This scarcity of the raw observational data calls for the discussions on the "lost cycle" just before the Dalton Minimum (*e.g.*, Usoskin *et al*., 2009; Zolotova and Ponyavin, 2011). On one hand, Usoskin *et al*. (2009) reconstructed butterfly diagram in 1790s to claim that occurrence of high solar latitude in 1793-1796 shows the start of "the lost cycle" in 1793. On the other hand, Zolotova and Ponyavin (2011) analyzed latitude-time diagram in 1784-1798 to claim that the local minimum in 1793 and "the lost cycle" were only a gap between impulses of the solar activity, possibly caused by the lack of data. Note that this "lost cycle" is under discussion in the viewpoint of contemporary cosmogenic radioisotope data as well (*e.g.* Karoff *et al*., 2015; Owens *et al*., 2015).

This sunspot drawing on 26 August 1793 is also valuable to be situated in the minimum candidate in 1793. In the viewpoint of $S_N$, this drawing supports the view of Vaquero *et al*. (2016) who confirmed relatively high number of sunspot groups. Nevertheless, unfortunately, this sunspot drawing is too isolated as shown in Table 1 and hence we cannot estimate the location of sunspots shown in this drawing. As long as we know, Iwahashi Zenbei left only one more drawing without explicit date (Figure 3). Other sunspot observations are not close enough to help estimating sunspot locations except for those by Schröter from 29 August 1793 (*e.g.* Vaquero *et al*., 2016). However, Schröter (S1794, p.265) himself provide only a short description on these sunspot observations as follows: "Not having for several days before, and likewise on the very day of the eclipse, noticed any spots on the disk of sun, three small ones only excepted, which were perceived on the 29th August". Considering that we do not have further descriptions or drawings, we cannot analyze the latitude of sunspots in this drawing on 26 August 1793 in the present stage. However, as attested by this sunspot drawing itself, we may have further unexamined sunspot drawings in the context of reconstructing $S_N$. Therefore, we believe further surveys on sunspot records and drawings around this date in comparison with contemporary data of cosmogenic radioisotopes (*e.g.* Karoff *et al*., 2015; Owens *et al*., 2015) may let us examine the distribution of sunspot latitudes to contribute our understanding on the discussion about "the lost cycle" around 1793. In this way, this drawing can contribute to the reconstruction of sunspot activity immediately before one of the grand minima.

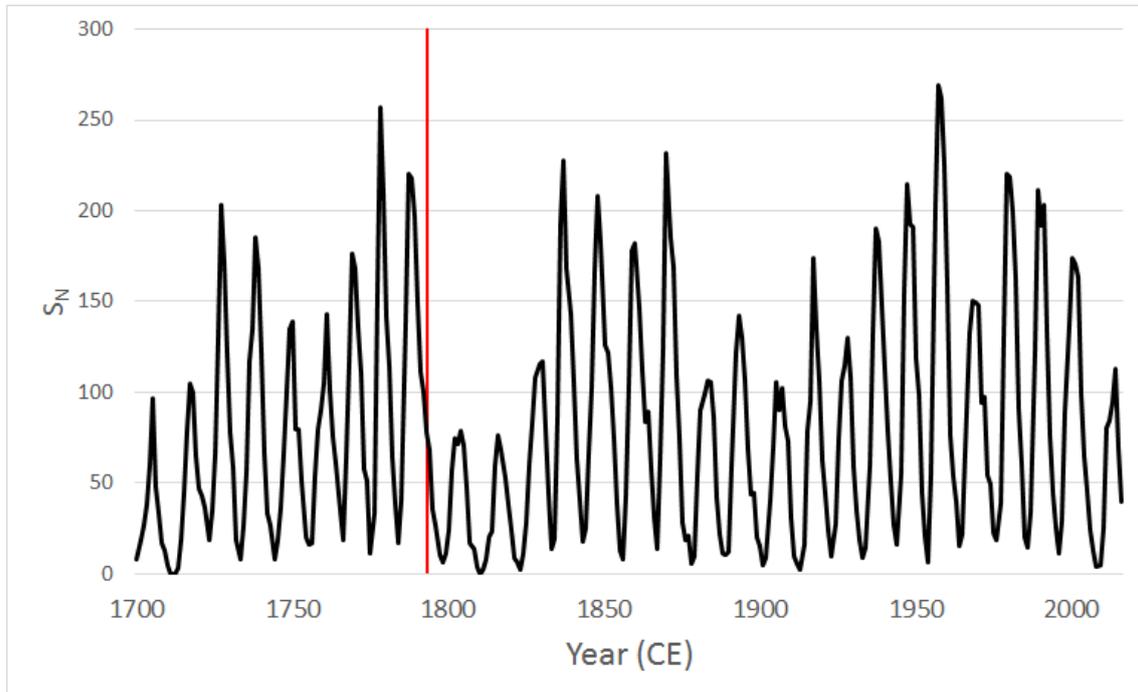

Figure 5a: Data context of Iwahashi Zenbei's sunspot drawing in the yearly $S_N$ reported by Clette *et al*. (2014).

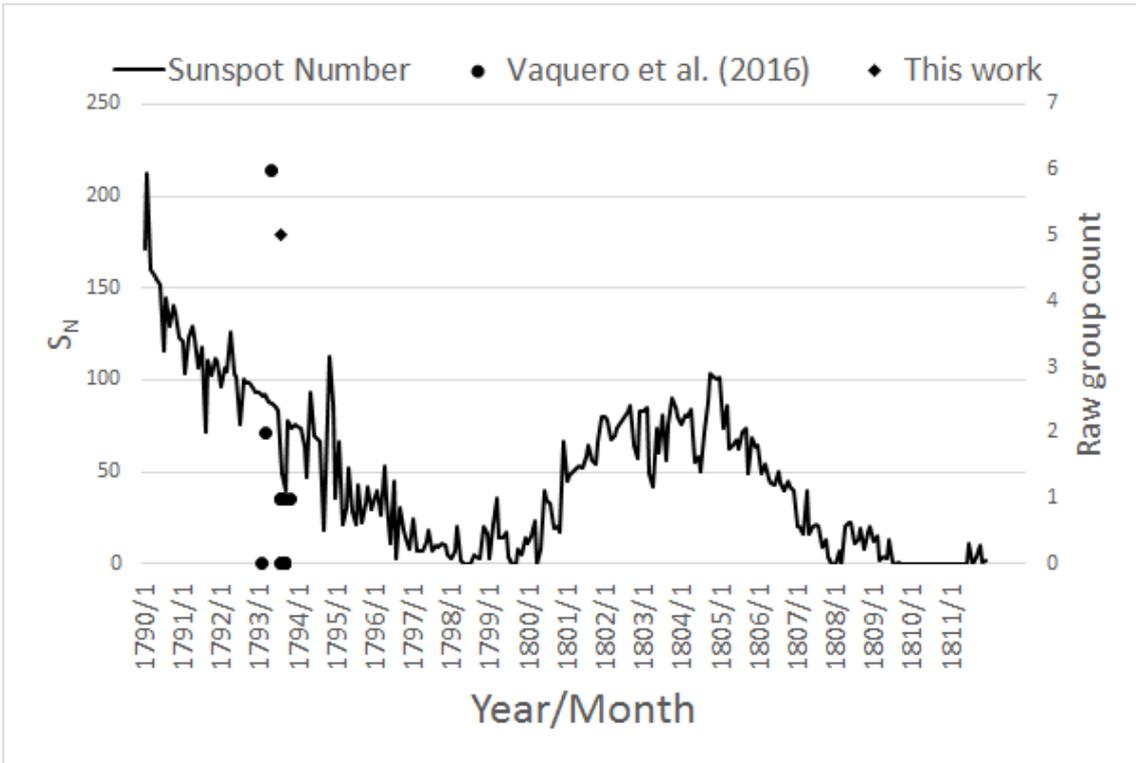

Figure 5b: Data context of Iwahashi Zenbei's sunspot drawing in the monthly $S_N$ reported by Clette *et al.* (2014).

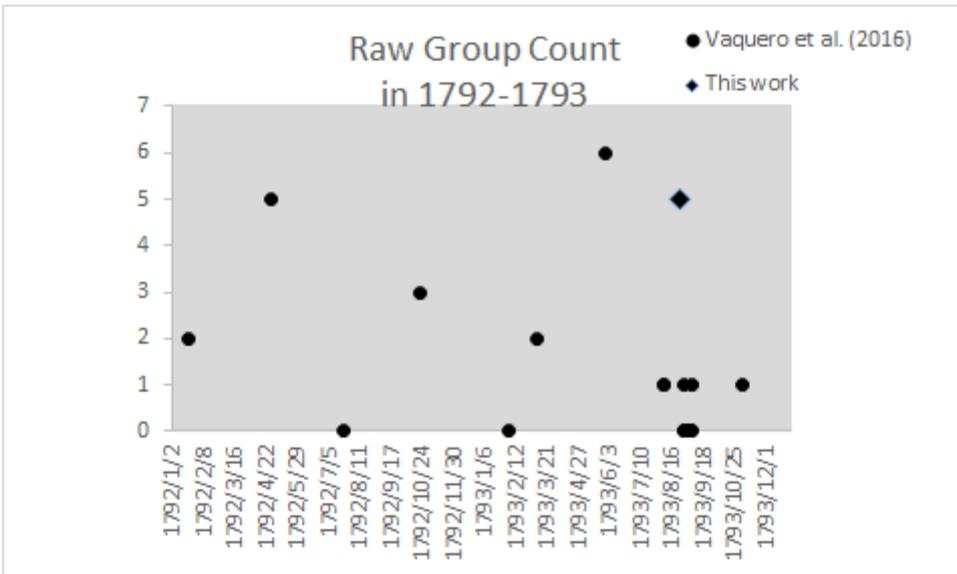

Figure 5c: Raw group count during 1792-1793 with Iwahashi Zenbei's observation

| Year | Month | Date | raw group count | Observer | Observational Site | Reference |
|------|-------|------|-----------------|----------|--------------------|-----------|
| 1793 | 2 | 4 | 0 | Hahn | Berlin | Vaquero et al. (2016) |
| 1793 | 3 | 9 | 2 | Staudach | Nuremberg | Vaquero et al. (2016) |
| 1793 | 5 | 28 | 6 | Huber | Basel | Vaquero et al. (2016) |
| 1793 | 8 | 5 | 1 | Staudach | Nuremberg | Vaquero et al. (2016) |
| 1793 | 8 | 6 | 1 | Staudach | Nuremberg | Vaquero et al. (2016) |
| 1793 | 8 | 26 | 5 | Iwahashi Zenbei | Fushimi | This Work |
| 1793 | 8 | 29 | 1 | Schroter | Lilienthal | Vaquero et al. (2016) |
| 1793 | 8 | 30 | 0 | Schroter | Lilienthal | Vaquero et al. (2016) |
| 1793 | 8 | 31 | 0 | Schroter | Lilienthal | Vaquero et al. (2016) |
| 1793 | 9 | 1 | 0 | Schroter | Lilienthal | Vaquero et al. (2016) |
| 1793 | 9 | 2 | 0 | Schroter | Lilienthal | Vaquero et al. (2016) |
| 1793 | 9 | 3 | 0 | Schroter | Lilienthal | Vaquero et al. (2016) |
| 1793 | 9 | 4 | 0 | Bode | Berlin | Vaquero et al. (2016) |
| 1793 | 9 | 4 | 0 | Schroter | Lilienthal | Vaquero et al. (2016) |
| 1793 | 9 | 4 | 0 | Staudach | Nuremberg | Vaquero et al. (2016) |
| 1793 | 9 | 5 | 0 | Bode | Berlin | Vaquero et al. (2016) |
| 1793 | 9 | 5 | 1 | Schroter | Lilienthal | Vaquero et al. (2016) |
| 1793 | 9 | 5 | 0 | Staudach | Nuremberg | Vaquero et al. (2016) |
| 1793 | 11 | 3 | 1 | Staudach | Nuremberg | Vaquero et al. (2016) |

Table 1: Data context of Iwahashi Zenbei's sunspot drawing in the raw group count reported by Vaquero et al. (2016).

## 4. Conclusion

In this article, we have shown two sunspot drawings by Iwahashi Zenbei and his companions. We have established the date of one of them to be 26 August 1793 but we could not find an explicit date for the other. Although, we do not have any further sunspot drawings by Iwahashi Zenbei, his sunspot observation can contribute to fill the not well-documented sunspot records in 1793, which is also located near the Dalton Minimum. This article also shows the further possibility of unexamined sunspot drawings in non-European countries before mid-19$^{th}$ century.

**Acknowledgement**

We acknowledge the support of the "UCHUGAKU" project of the Unit of Synergetic Studies for Space, the Exploratory and Mission Research Projects of the Research Institute for Sustainable Humanosphere (PI: H. Isobe), and SPIRITS 2017 (PI: Y. Kano) of Kyoto University. This work was also encouraged by a Grant-in-Aid from the Ministry of Education, Culture, Sports, Science and Technology of Japan, Grant Number JP15H05816 (PI: S. Yoden), JP16H03955 (PI: K. Shibata), JP16K17671 (PI: S. Toriumi), and JP15H05814 (PI: K. Ichimoto), and Grant-in-Aid for JSPS

Research Fellow JP17J06954 (PI: H. Hayakawa). We especially thank the National Institute for Japanese Literature, Tsu City Library, the International Research Center for Japanese Studies, and the Library of the National Astronomical Observatory of Japan for granting permission for research and reproductions of the original manuscripts and woodprints. Note that the latter's ongoing digitalization project is supported by the Project to Build an International Collaborative Research Network for Pre-modern Japanese Texts (NIJL-NW project), to which KI also thanks. We also thank SIDC for providing data on group sunspot number and raw group count of sunspots, Dr. H. Isobe for preliminary review on our manuscript, and Mr. K. Wada for providing valuable advice on identification of observational sites of Iwahashi Zenbei.

**Disclosure of Potential Conflicts of Interest**

The authors declare that they have no conflicts of interest.

**Appendix 1: References of Historical Documents**

BK-J: Tachibana Nankei (橘南谿), *Bouenkyo Kanshoyoki* (望遠鏡観諸曜記), MS 463 in the Library of the International Research Center for Japanese Studies [a manuscript in Japanese]